# Data Mining Model for the Data Retrieval from Central Server Configuration

Srivatsan Sridharan[1], Kausal Malladi and Yamini Muralitharan[2]

[1] Department of Computer Science, IIIT - Bangalore, India.
{vatsansridharan,kausalmalladi}@gmail.com

[2] Department of Software Engineering, IIIT - Bangalore, India.
{yams25}@gmail.com

*Abstract.*

*A server, which is to keep track of heavy document traffic, is unable to filter the documents that are most relevant and updated for continuous text search queries. This paper focuses on handling continuous text extraction sustaining high document traffic. The main objective is to retrieve recent updated documents that are most relevant to the query by applying sliding window technique. Our solution indexes the streamed documents in the main memory with structure based on the principles of inverted file, and processes document arrival and expiration events with incremental threshold-based method. It also ensures elimination of duplicate document retrieval using unsupervised duplicate detection. The documents are ranked based on user feedback and given higher priority for retrieval.*

## 1. Introduction

Data intensive applications such as electronic mail, news feed, telecommunication management, automation of business reporting etc raise the need for a continuous text search and monitoring model. In this model the documents arrive at the monitoring server as in the form of a stream. Each query Q continuously retrieves, from a sliding window of the most recent documents, the k that is most similar to a fixed set of search terms. *Sliding window.* This window reflects the interest of the users in the newest available documents. It can be defined in two alternative ways. They are a) *count-based window* contains the N most recent documents for some constant number N, b) *time-based window* contains only documents that arrived within last N time units. Thus, although a document which may be relevant to a query, it is ignored, because it may not satisfy the time and count constraints of the user. *Incremental threshold method.* The quintessence of the algorithm is to employ threshold-based techniques to derive the initial result for a query, and then continue to update the threshold to reflect document arrivals and expirations. At its core lies a memory-based index similar to the conventional inverted file, complimented with fast updated techniques. *MapReduce technique.* MapReduce is a powerful platform for large scale data processing. This technique involves two steps namely a) *map step:* The master node takes the input, partitions it up into smaller sub-problems, and distributes them to worker nodes. A worker node may do this again in turn, leading to a multi-level structure. The worker node processes the smaller problem, and passes the answer back to its master node, b) *reduce step:* The master node then collects the answers to all the sub-problems and combines them in some way to form the output – the answer to the problem it was originally trying to solve. *Unsupervised duplicate detection.* [3] The problem of identifying objects in databases that refer to the same real world entity, is known, among others, as duplicate detection or record linkage. Here this method is used to identify documents that are all alike and prevent them from being prepared in the result set. Our paper also focuses on ranking the documents based on user feedback. The user is allowed to give feedback for each document that has been retrieved. This feedback is used to rank the document and hence increase the probability of the document to appear in the sliding window.

*Visual Web Spider* is a fully automated, multi-threaded web crawler that allows us to index and collect specific web pages on the Internet. Once installed, it enables us to browse the Web in an automated manner, indexing pages that contain specific keywords and phrases and exporting the indexed data to a database on our local computer in the format of our choice. We want to collect website links to build our own specialized web directory. We can configure **Visual Web Spider** automatically. This program's friendly, wizard-driven interface lets us customize our search in a step-by-step manner. To index relevant web pages, just follow this simple sequence of steps. After opening the wizard, enter the starting web page URL or let the program generate URL links based on specific keywords or phrases. Then set the crawling rules and depth according to your search strategy. Finally, specify the data you want to index and your project filename. That's pretty much it. Clicking on 'Start' sets the crawler to work. Crawling is fast, thanks to multi-threading that allows up to 50 simultaneous threads. Another nice touch is that Visual Web Spider can index the content of any HTML tag such as: page title (TITLE tag), page text (BODY tag), HTML code (HTML tag), header text (H1-H6 tags), bold text (B tags), anchor text (A tags), alt text (IMG tag, ALT attribute), keywords, description (META tags) and others. This program can also list each page size and last modified date. Once the web pages have been indexed, Visual Web Spider can export the indexed data to any of the following formats: Microsoft Access, Excel (CSV), TXT, HTML, and MySQL script.

## 1.1. Key Features

A Personal, Customizable Web crawler. Crawling rules. Multi-threaded technology (up to 50 threads). Support for the robots exclusion protocol/standard (Robots.txt file and Robots META tags);Index the contents of any HTML tag. Indexing rules; Export the indexed data into Microsoft Access database, TEXT file, Excel file (CSV), HTML file, MySQL script file; Start crawling from a list of the URLs specified by user; Start crawling using keywords and phrases; Store web pages and media files on your local disk;  Auto-resolve URL of redirected links; Detect broken links; Filter the indexed data;

## 2. Existing System

Drawbacks of the existing servers that tend to handle the heavy document traffic are: Cannot efficiently monitor the data stream that has highly dynamic document traffic. The server alone does the processing hence it involves large amount of time consumption. In case of continuous text search queries and extraction every time the entire document set has to be scanned in order to find the relevant documents. There is no confirmation that duplicate documents are not retrieved for the given query. A large amount of documents cannot be stored in the main memory as it involves large amount of CPU cost. *Naïve solution:* The most straightforward approach to evaluate the continuous queries defined above is to scan the entire window contents D after every update or in fixed time intervals, compute all the document scores, and report the top-k documents. This method incurs high processing costs due to the need for frequent re computations from scratch.

# 3. Proposed System

## 3.1. Problem Formulation

In our model, a stream of documents flows into a central server. The user registers text queries at the server, which is then responsible for continuously monitoring/reporting their results. As in most stream processing systems, we store all the data in main memory in order to cope with frequent updates, and design our methods with the primary goal of minimizing the CPU cost. Moreover it is necessary to reduce the work load of the monitoring server.

## 3.2. Proposed Solution

In our solution we use the MapReduce technique in order to reduce the work load of the central server, where the server acts as the master node, which splits up the processing task to several worker nodes. The number of worker nodes, which have been assigned the processing task, depends on the nature of query that has been put up by the user. Here the master node, upon receiving a query from the user, assigns the workers to find the relevant result query set and return the solution to the master node. The master node, after receiving the partial solutions from the workers, integrates the results to produce the final result set for the given query. This can be viewed schematically in the following *Fig.1*. Each worker/slave node is responsible uses the *incremental threshold algorithm* for computing the result set of k relevant and recent documents for the given query. The overall system architecture can be viewed as in the following *Fig.2*

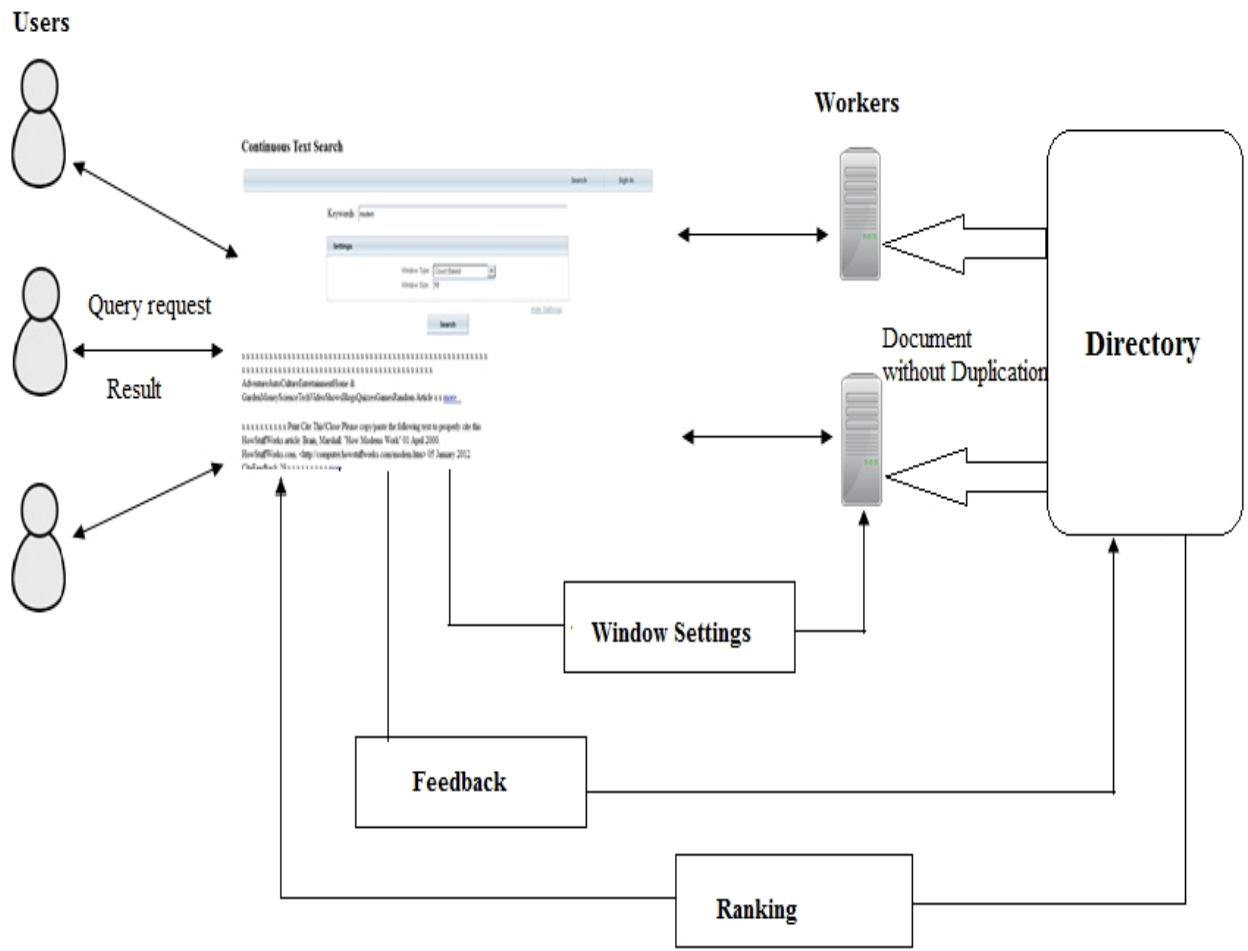

Figure. 1. System Architecture for the proposed Data Retrieval Model.

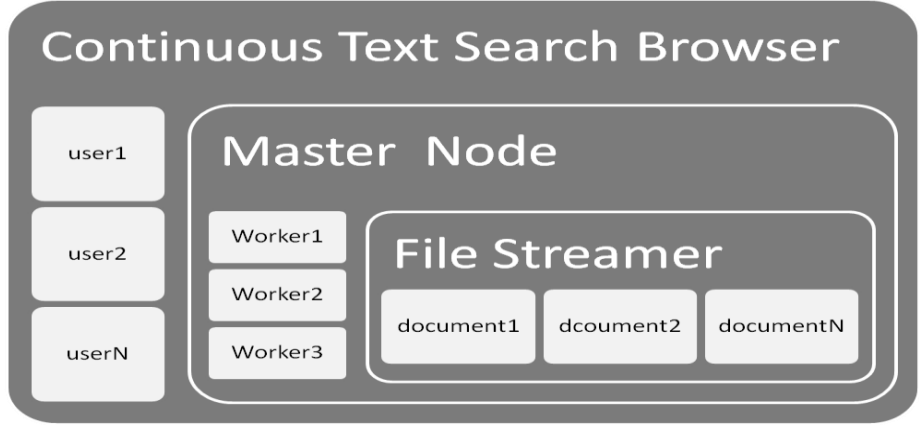

**Fig. 2.** A data Retrieval system using MapReduce.

Each element of the input stream comprises of a document d, a unique document identifier, the document arrival time, a composition list. The composition list contains one (t, wdt) pair for each term t belonging to T in the document and wdt is the frequency of the term in the document d. The notations in this model are as follows in Fig 3.

| Symbol | Description |
|---|---|
| $T$ | dictionary of all possible terms of interest |
| $\lambda$ | document arrival rate |
| $N$ | sliding window size |
| $D$ | set of valid documents (in the current window) |
| $Q$ | a continuous text query |
| $q$ | number of queries |
| $n$ | number of query terms |
| $R$ | query result list |
| $k$ | number of result documents |
| $S(d|Q)$ | similarity score of document $d$ w.r.t. $Q$ |
| $S_k$ | score of the $k$-th document in $R$ |
| $\tau$ | global/influence threshold |
| $\theta_{Q,t}$ | local threshold of query $Q$ for term $t$ |

**Figure. 3.** A Detailed list of the notations.

The worker node maintains an inverted index for each term t in the document. With the inverted index, a query Q is processed as follows: the inverted lists for the terms t belonging to Q are scanned and the partial wdt scores of each encountered document d are accumulated to produce S(d/Q). The documents with the highest scores at the end are returned as the result.

### 3.3. Incremental Threshold Algorithm

Fig.3 represents the data structures that have been used in this system. The valid documents D are stored in a single list, shown at the bottom of the figure. Each element of the list holds the stream of information of document (identifier, text content, composition list, arrival time). D contains the most recent documents for both count-based and time-based windows. Since documents expire in first-in-first-out manner, D is maintained efficiently by inserting arriving documents at the end of the list and deleting expiring ones from its head. On the top of the list of valid documents we build an inverted index. The structure at the top of the figure is the dictionary of search terms. It is an array that contains an entry for each term t belonging to T. The dictionary entry for t stores a pointer to the corresponding inverted list $L_t$. $L_t$ holds an impact entry for each document d that contains t, together with a pointer to d's full information in the document list. When a document d arrives, an impact entry (d, wdt) (derived from d's composition list) is inserted into the inverted list of each term t that appears in d. Likewise, the impact entries of an entries of an expiring document are removed from the respective inverted lists. To keep the inverted lists sorted on wdt while supporting fast (logarithmic) insertions and deletions.

*Initial Top-k Search:* When a query is first submitted to the system, its top-k result is computed using the initial search module. The process is an adaptation of the threshold algorithm. Here, the inverted lists $L_t$ of the query terms play the role of the sorted attribute lists. Unlike the original threshold algorithm, however we do not probe the lists in a round robin fashion. Since the similarity function associates different weights wQt with the query specifically, inspired by [4], we probe the list $L_t$ with the highest $c_t$=wQt.wd$_{nxt}$t value, where $d_{nxt}$ is the next document in $L_t$. The global threshold $g_t$, a notion used identically to the original algorithm, is the sum of $c_t$ values for all the terms in Q. Consider query $Q_1$ with search string "red rose" and k=2. Let term $t_{20}$="red" and $t_{11}$="rose". First the server identifies the inverted lists $L_{11}$ and $L_{20}$ (using the dictionary hash table), and computes the values $c_{11}$=wQ$_1$t$_{11}$.wd$_7$t$_{11}$ and $c_{20}$=wQ$_1$t$_{20}$.wd$_6$t$_{20}$. In iteration 1, since $c_{20}$ is larger, the first entry of $L_{20}$ is popped; the similarity score of the corresponding document, $d_6$, is computed by accessing its composition list in D and inserted into the tentative R. $c_{20}$ is then updated to impact entry which is above local threshold, but we would still include it in R as unverified entry. The algorithm is as follows,

*Algorithm Incremental Threshold with Duplicate Detection (Arriving $d_{ins}$, Expiring $d_{del}$)*

1: Insert document $d_{ins}$ into D (the system document list)

2: for all terms t in the composition list of $d_{ins}$ do

3: for all documents in $L_t$

4: for all terms t in $d_{ins}$

5: Compute unique ($d_{ins}$)

6: wd$_{ins}$t != wd$_{nxt}$t

7:  Insert the impact entry of $d_{ins}$ into $L_t$

8:  Probe the threshold tree of $L_t$

9:  for all queries Q where wd$_{ins}$t >= localThreshold do

10:   if Q has not been considered for $d_{ins}$ in another $L_t$ then

11:    Compute S ($d_{ins}$/Q)

12:    Insert $d_{ins}$ into R

13:  if S($d_{ins}$/Q)>= old $S_k$ then

14:  Update $S_k$ (since $d_{ins}$ enters the top-k result)

15:  Keep rolling up local thresholds while r <= $S_k$

16:  Set new τ as influence threshold for Q

17:    Update local thresholds of Q

18:    Delete document $d_{del}$ from D (the system document list)

19:    for all terms t in the composition list of $d_{del}$ do

20:         Delete the impact entry of $d_{del}$ from $L_t$

21:         Probe the threshold tree of $L_t$

22:         for all queries Q where $wd_{del}t$ >= localThreshold do

23:         if Q has not been considered    for $d_{del}$ in another $L_t$ then

24:    Delete $d_{del}$ from R

25:    if $S(d_{del}/Q)$ >= old $S_k$ then

26:       Resume top-k search from local thresholds

27:    Set new τ as influence threshold for Q

28: Update local thresholds of Q

After constructing the initial result set R using the above algorithm, only the documents that have a score higher than or equal to the influence threshold t(tow) are verified. The main key point is that no duplicate documents from the part of the result set R. This is ensured using unsupervised duplication detection. The idea of unsupervised learning for duplicate detection has its roots in the probabilistic model proposed by Fellegi and Sunter. When there is no training data to compute the probability estimates, it is possible to use variations of the Expectation Maximization algorithm to identify appropriate clusters in the data.

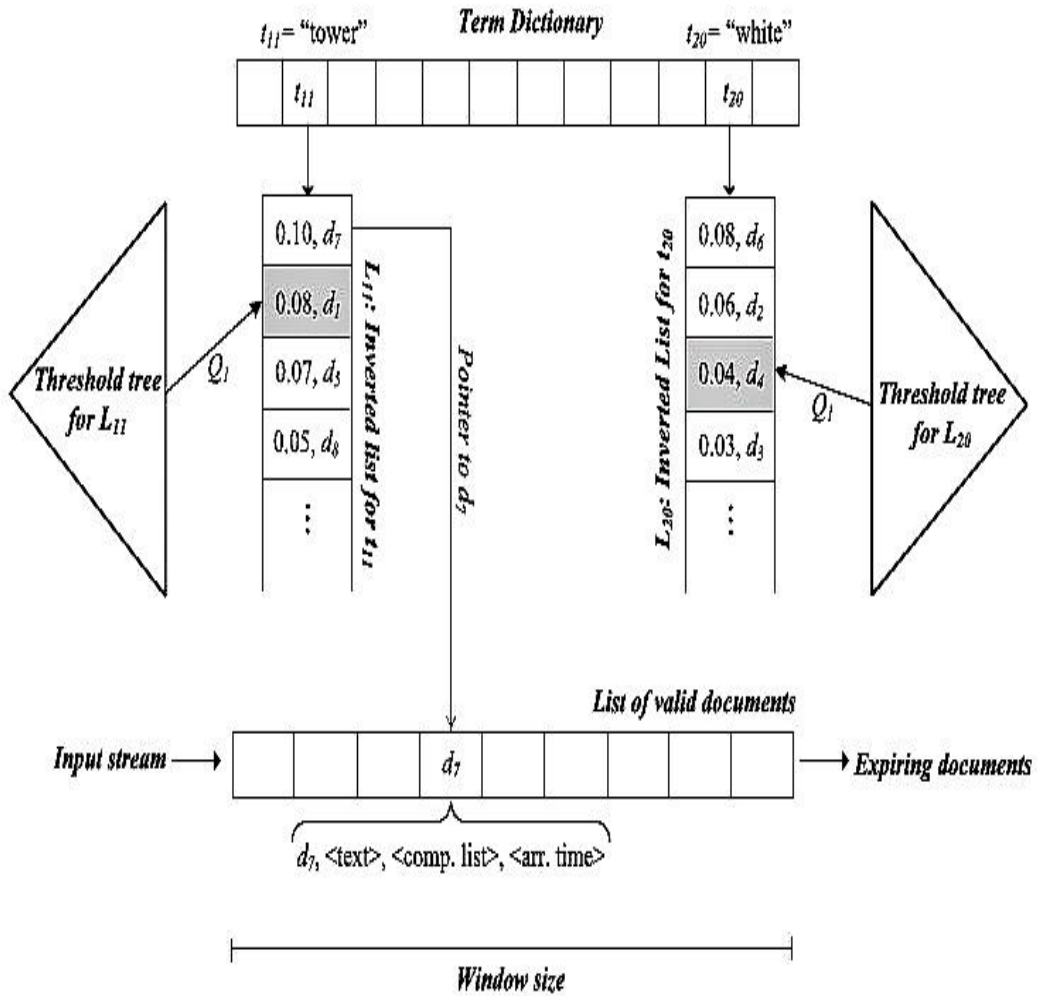

**Figure. 4.** Data Structures used for Incremental Threshold Algorithm.

## 4. Performance and Implementation Issues

In Figure.5 we empirically compare Incremental Threshold Algorithm (ITA) against Naïve. We enhance Naïve with the technique of, which retrieves the top-$k_{max}$ documents (for an analytically derived $k_{max}$ that is larger than k) whenever the result is computed from scratch, in order to reduce the frequency of subsequent recomputations. We form a document stream, which comprises 172,961 articles. After standard stopword removal, the dictionary contains 181,978 terms. The documents are streamed into the monitoring system, following a Poisson process with a mean arrival rate of 200 documents/second. We generate 1,000 queries with k = 10 and terms selected randomly from the dictionary. We use a count-based window; the results for a time-based one are similar. First, we investigate the effect of the number of search terms n on the performance of ITA and Naïve. In Figure 5(a), we set the window size to 1,000 documents, and vary n from 4 to 40. We measure the average processing time, i.e., the elapsed time between the arrival of a new document (which additionally causes the expiration of an existing one) and the point where all the query results are updated accordingly. The measurements are in milliseconds and are plotted in logarithmic scale. With more search terms (i.e., larger n), an arriving/expiring document has a higher chance of sharing common terms with the queries. This leads to an increase in the number of queries that need updating, and thus to a longer processing time. ITA is about 10 times faster than Naïve for queries comprising 4 search terms, and 6 times faster for 40- term queries.

| Querylength | Processing time for naïve in msec | Processing time for ITA in msec |
|---|---|---|
| 3 | 0.9 | 0.07 |
| 8 | 1.5 | 0.2 |
| 13 | 3 | 0.4 |
| 18 | 8 | 0.7 |
| 23 | 13 | 1.0 |

**Table 1.** Data for Query Length and processing Time.

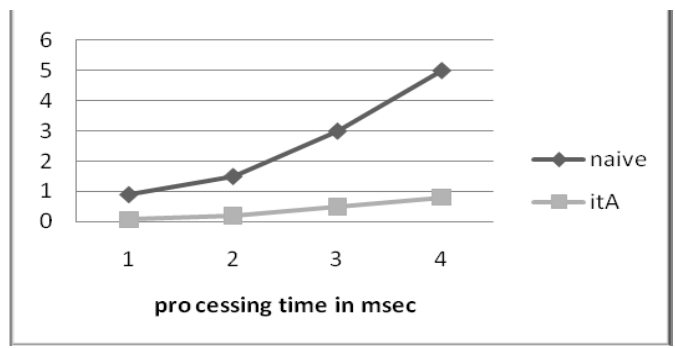

**Figure. 5a.** Comparison of Query Length and processing Time.

In Figure 5(b), we study the effect of the sliding window size N. We set the query length to 10 terms, and vary N from 10 to 100,000 documents. A larger window holds more valid documents in the system. For Naïve this imposes a higher cost whenever the result needs to be recomputed, because it scans the entire D. For ITA, the inverted lists grow longer, leading to higher index update cost and slower arrival/expiration handling. ITA is

13 times faster than Naïve for a window size of 10, and 18 times faster when the sliding window comprises 10,000 documents. Note that the last measurement for Naïve is missing, because for N > 10000 the CPU utilization approaches 100% and the system becomes unstable. The above experiments, as well as others omitted due to lack of space, verify the general superiority of ITA over Naïve.

| Window size | Processing time for naïve in msec | Processing time for ITA in msec |
|---|---|---|
| 5 | 10 | 5 |
| 13 | 20 | 7 |
| 25 | 30 | 13 |
| 30 | 40 | 15 |

**Table 1.** Data for Window Size and processing Time.

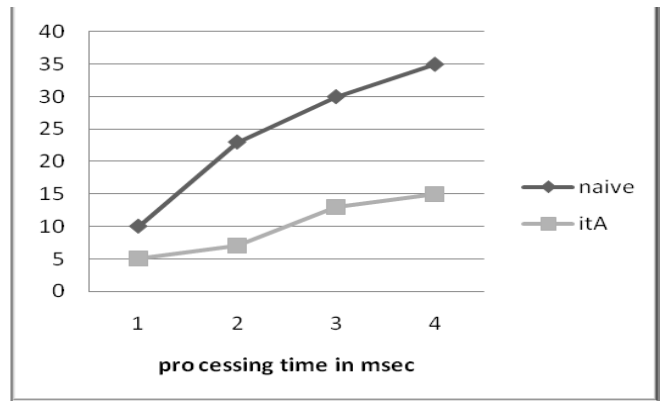

**Figure. 5b.** Comparison of Window Size and processing Time.

## 5. Conclusion

In this paper, we study the processing of continuous text queries over document streams. These queries define a set of search terms, and request continual monitoring of a ranked list of recent documents that are most similar to those terms. The problem arises in a variety of text monitoring applications, e.g., email and news tracking. To the best of our knowledge, this is the first attempt to address this important problem. Currently, our study focuses on plain text documents. A challenging direction for future work is to extend our methodology to documents tagged with metadata and documents with a hyperlink structure, as well as to specialized scoring mechanisms that may apply in these settings.

### ACKNOWLEDGEMENTS

The authors would like to thank Prof. R. Parvatham Assistant Professor, Sri SaiRam Engineering College for her continuous support in implementing this work.

**Authors.**

**Srivatsan Sridharan.** Srivatsan is a young academician who is highly passionate to learn and adapt himself to newer technologies. Completed his Bachelor of Engineering from Sri SaiRam Engineering College affiliated to Anna University - Chennai (Ranked 26[th] among all the Affiliated Institutions in the Dept. of Computer Science), he is currently pursuing his M.Tech from IIIT - Bangalore. He has been awarded thrice the Best Paper Award, One from ICCCAN 2013, another from IEEE - ICRDPET 2013 and the other from ICEAT -2012 Conference. He has also applied for Four Patents in Indian Patents office Located at Chennai, Tamilnadu, India, two with Kausal Malladi, one with Yamini Muralitharan and one individually.

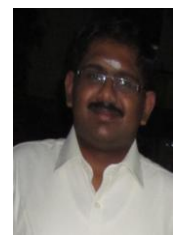

**Kausal Malladi**. A B.Tech. Graduate in Information Technology, result-driven engineer with a traction towards open source, visionary entrepreneur, currently pursuing Masters in Computer Science from International Institute of Information Technology, Bangalore (IIIT-B). Passionate about technology, looking forward to solve few of the existing problems in the arena of computers by extending expertise in Operating Systems, Data Management and Theoretical Computer Science. He has applied for two patents jointly with Srivatsan Sridharan.

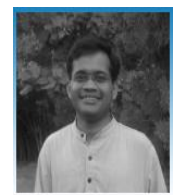

**Yamini Muralitharan.** Yamini is a young academician who is much passionate towards research in Software Engineering and Testing. She completed her Bachelors from College of Engineering - Guindy - Anna University, Tamilnadu. She is currently pursuing her Masters from IIIT - Bangalore. She has also applied for One Patent jointly with Srivatsan Sridharan at Patents Office - Chennai.

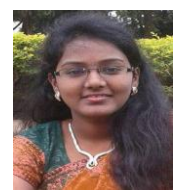